
\magnification=1200
\input amssym.def
\input amssym.tex

\font\boldtitlefont=cmb10 scaled\magstep2
\font\sectionfont=cmb10 scaled\magstep1

\def\nspace{\lineskip=1pt\baselineskip=12pt\lineskiplimit=0pt}

\def\half{{1\over 2}}
\def\references#1{\bigskip\par
\noindent{\sectionfont References}\medskip
     \parindent=#1pt\nspace}
\def\ref#1{\par\smallskip\hang\indent\llap{\hbox to\parindent
     {#1\hfil\enspace}}\ignorespaces}
\def\half{{1\over2}}

\def\eps{{\varepsilon}}

\def\ln{\rm ln}

\footline={\hfill}

\parskip=\smallskipamount
\parindent=25pt

\centerline{\boldtitlefont Conformal Field Theory and}

\smallskip
\centerline{\boldtitlefont Hyperbolic Geometry}

\bigskip
\centerline{{\bf P. Kleban$^{1,2,3}$ and
I. Vassileva$^{2,4}$}}

\bigskip
\centerline{${}^1$ School of Mathematics}
\centerline{Institute for Advanced Study}
\centerline{Princeton, New Jersey \ 08540}

\medskip
\centerline{${}^2$LASST and ${}^3$Department of}
\centerline{Physics \& Astronomy}
\centerline{University of Maine}
\centerline{Orono, Maine \ 04469\footnote{*}%
{Present and permanent address. \
e-mail: kleban@maine.maine.edu}}

\medskip
\centerline{${}^4$Department of Mathematics}
\centerline{Univeristy of Massachusetts}
\centerline{Amherst, Massachusetts \ 01003}

\vskip4pc
\centerline{\sectionfont Abstract}

{\narrower{\narrower\bigskip
\noindent
We examine the correspondence between the conformal field
theory of boundary operators and two-dimensional hyperbolic
geometry.
by consideration of domain boundaries in two-dimensional
critical systems, and the invariance of the  hyperbolic
length, we motivate a reformulation of the basic equation of
conformal covariance.
The scale factors gain a new, physical interpretation.
We exhibit a fully factored form for the three-point
function.
A doubly-infinite discrete series of central charges with
limit $c=-2$ is discovered.
A correspondence between the anomalous dimension and the
angle of certain hyperbolic figures emerges.\smallskip}}

\vfill\eject
\footline={\hss\tenrm\folio\hss}
\pageno=1

In this letter, we establish several connections between the
conformal field theory of boundary operators and
two-dimensional hyperbolic geometry.
First, by consideration of domain boundaries in
two-dimensional critical systems, and the invriance of the
hyperbolic length, we motivate a reformulation of the basic
equation of conformal covariance.
The scale factors gain a new, physical interpretation.
They operate to keep the distance from the end of the domain
boundary to the boundary of the system fixed.
We also point out that for any geometry conformally
equivalent to the half-plane, domain boundaries in
two-dimensional critical systems follow hyperbolic
geodesics.
Their energy per unit hyperbolic length is finite.
Motivated by these results, we next exhibit a completely
factored form for the three-point function (and the
prefactor of the four-point function).
Here, a connection between the anomalous dimension of a
primary operator and the angle of a hyperbolic figure
appears.
Finally, we impose the condition that the Schwarz function
defined by a four-point function of opertors degenerate at
level two correspond to a hyperbolic tiling, or {\it
tessellation}.
this leads to a new, doubly-infinite discrete set of minimal
models.
The angle-dimension correspondence is again encountered.

To begin, we establish a connection between conformal field
theory and hyperbolic geometry in the language of the theory
of phase transitions.
However, it should be emphasized that our results are
generally valid,a nd not dependent on this particular
realization of the theory.

As demonstrated elsehwere [1], a domain boundary in the
upper half-plane is created by boundary operators $\phi(x)$
[1--5] located at its endpoints on the real axis.
These operators act to change the boundary condition along
the edge of the system [3], the real axis.
Boundary operators may also be defined by letting bulk
operator in the system with a boundary approach the boundary,
and making use of the bulk-boundary oeprtor product
expansion [1,4].

Although a domain boundary at a critical point exhibits
large fluctuations, and has an energy that is not
proportional to its length, it is a well-defined object.
Conformal invariance implies universality, which allows one
to study it in general.
The (extra free) energy of such a boundary is
$$
F=-\ln\left<\phi(x_1)\phi(x_2)\right>,\eqno{(1)}
$$
as described in [1].
For completeness, we note tht Equation (1) ignores both
universal [3,6] and non-universal constants independent of
$x_1$, $x_2$.
The former are associated with the boundary states on the
real axis, while the latter arise in computing the free
energy of the boundary of any real system of statistical
mechanical model.

Evaluating the correlation funciton, we find [1]
$$
F=2\Delta \ln\vert x_1-x_2\vert,\eqno{(2)}
$$
where $\Delta$ is the critical dimension of $\phi$.
Now, Equation (1) also gives the domain boundary free energy
in any geometry conformally equivlaent to the half-plane, if
we evaluate the correlation function in the new geometry.
This is done by making use of the basic equation of
conformal covariance of correlation functions [7], as
applied to boundary operators,
$$
\left<\phi_1(x_1)\phi_2(x_2)\ldots\right>=
\vert w(x_1)\vert^{\Delta_1}\vert w(x_2)\vert^{\Delta_2}
\left<\phi_1(w_1)\phi_2(w_2)\ldots\right>.
\eqno{(3)}
$$
Here $w=w(z)$ is an arbitrary conformal transformation, with
$w_i=w(x_1)$.

Now, the domain boundary itself, in the half-plane, is a
half circle betweent he points $x_1$ and $x_2$.
This follows by considering a single change of boundary
conditions, at the origin, say.
By symmetry, the corresponding boundary lies along the
$y$-axis.
A projective transformation brings the endpoints to $x_1$
and $x_2$, and the straight line becomes a half circle.
Such a half circle is precisely a geodesic of the
hyperbolic, or Poincar\'e, metric $ds_h^2={1\over y^2}ds^2$
(for $y>0$) [8].
The tendency of the domain boundary to avoid the real axis
corresponds to the divergence of the hyperbolic length as
$y->0$.

Next consider the hyperbolic geodesic between the points
$z_1=x_1+\eps_1$, and $z_2=x_2+\eps_2$.
Its hyperbolic length follows from standard results [8]
$$
l_k=2\ln\vert x_1-x_2\vert-\ln\,\eps_1\eps_2,\eqno{(4)}
$$
and diverges as $z_1$ or $z_2$ approaches the real axis.
On the other hand, it is natural to define the domain
boundary to begin and end at a fintie but small (Euclidean)
distance $\eps$ above the real axis, where $\eps$ will be on
the order of the lattice spacing in any physical model.
Then, up to additive constants, the boundary energy is
proportional to the hyperbolic length,
$$
F=\Delta(l_h+\ln\,\eps^2).\eqno{(5)}
$$
Note that factors giving rise to the $\ln\,\eps^2$ term in
Equation (5) will appear naturally in any lattice
calculation of $F$ in a spin model, through the
normalization of the conformal operators [9].

Next consider any other geometry that can be mapped to the
half plane by a conformal transformation, for instance an
infinite strip (with edges) of width $L$, $w={L\over\pi} \ln\,
z$.
Under any such transformation, the hyperbolic length is
invariant.
The hyperbolic metric in the new geometry is induced by the
transformation.
In the strip, for instance,
$g={(\pi/L)^2\over \sin^2(\pi\nu/L)}$.
Although the hyperbolic length of the boundary remains
fixed, the transformation changes the distance between each
endpoint and the edge of the system, by an amount
proportional to the scale factor $\vert w(x)\vert$.
For the transformed theory to represent a physical domain
boundary in the new geometry, one must readjust its endpoints
to be at distance $\eps$ from the edges.
Using the invariance mentioned and Equation (1) then leads
directly to Equatiion (3), which is the basis of conformal
field theory.
The scale factors appear in (to our knowledge) an entirely
new interpretation -- they are necessary to insure that, in
the new geometry, the boundary begins and ends in the
appropriate place.

The argument of the preceding paragraph is not quite
complete.
The results described provide a hyperbolic interpretation
for an arbitrary two-point correlation function of boundary
operators.
Specification of the full theory involves higher point
correlation functions.
The new element that appears is their dependence on
cross-ratios [10]
$$
C={(x_1-x_2)(x_3-x_4)\over(x_2-x_4)(x_1-x_3)}.\eqno{(6)}
$$
However, if we consider the points $z_i=x_i+i\eps$, it is
easy to see that
$$
C=\exp\half\{l_h(1,2)+l_h(3,4)-l_h(2,4)-l_h(1,3)\}.
\eqno{(7)}
$$
Now, as mentioned, $l_h$ is invariant under conformal
transformations.
The combination that appears in Equation (7) is, in
addition, unaffected by the scale factors.
Equation (3) thus follows immediately.
It should be recognized that when more than one oeprator is
included in the correlation function, a weighted sum of
hyperbolic lengths will appear in the logarithm of the
correlation function, with the weighting depending on the
dimensions of the operators.

Note that in many cases four (and higher) point correlation
functions can be interpreted in terms of domain boundary
energies, including interactions [1].
In the case of critical percolation, an alternative
interpretation of boundary operator correlation functions
in terms of the probabilities of events is possible.
This view has been exploited for the description of crossing
probabilities in finite geometries [11--13].

The boundary operator theory is in general completely
equivalent to the corresponding bulk conformal theory
[4,5,14], in the sense that, for a given central charge, the
spectrum of primary operators is the same.
However, a given  operator will generally play a different
role than in the bulk.

It should be emphasized that the systems to which the theory
applies are defined in flat space.
The hyperbolic geometry, which describes a space of constant
negative curvature, arises naturally from the mathematics,
without having been put in at the start, by contrast to
other treatments of field theories [15] and statistical
systems [16] defined on hyperbolic spaces.
 From a mathematical point of view, this is not completely
unexpected.
To represent a physical theory, the metric must be
equivalent at equivalent points in the half-plane.
Only two metrics satisfy this condition -- Euclidean
and hyperbolic.
Of course this only means that the hyperbolic metric {\it
can} occur, not that it necessarily will.

We pause to consider some implications of our results thus
far.
We have shown that the invariance of the hyperbolic length
and natural physical requirements lead to a new derivation
of the basic equation of conformal covariance, including a
new interpretation of the scale factors.
These considerations suggest that there is a fundamental
mathematical connection between conformal field theory in
the half-plane and hyperbolic geometry.
Also, we have demonstrated that the energy of a domain
boundary in any geometry conformally equivalent to the
half-plane (e.g., Equation (5) of [1]) appears as a line
integral along a hyperbolic geodesic.
(In fact, it was precisely the search for a representation
of the energy as a line integral that led to these results.)
 From a physical viewpoint, this is very surprising.
At the critical point the boundary is strongly fluctuating
-- for instance, in a strip of width $L$, the boundary's
width will be of the same order.
There is absolutely no reason to expect that one of its
intrinsic properties can be described as a line integral.
A general picture of domain boundaries as independent,
weakly interacting objects was established in [1].
The fact that the energy of a single boundary is
proportional to its hyperbolic length, as described above,
illuminates its nature further.
More specifically, it indicates that a boundary, despite its
large fluctuations, is in some sense additive.

Next consider an arbitrary three-point function.
This has the form [7]
$$
\left<\phi_l(x_1)\phi_m(x_2)\phi_n(x_3)\right>=
{C_{lmn}\over x_{21}^{\Delta_l+\Delta_m-\Delta_n}
x_{32}^{\Delta_m+\Delta_n-\Delta_l}x_{31}^{\Delta_l+
\Delta_n-\Delta_m}},\eqno{(8)}
$$
where $C_{lmn}$ is an operator product expansion
coefficient, $x_{ji}=x_j-x_i$, and we have taken
$x_1<x_2<x_3$.
Now consider the hyperbolic triangle defined by the three
points $z_i=x_i+i\eps$, $i=1,2,3$.
Using the cosine law for hyperbolic triangles [8], it is
then straightforward to show that the angles at points
$x_1$, $x_2$, $x_3$ are of the form $a\eps$, $b\eps$,
$c\eps$, respectively, with $a=2$ $x_{32}/x_{21}x_{31}$,
etc.
It follows that
$$
\left<\phi_l(x_1)\phi_m(x_2)\phi_n(x_3)\right>
=C_{lmn}\left({a\over 2}\right)^{\Delta_l}
\left({b\over 2}\right)^{\Delta_m}
\left({c\over 2}\right)^{\Delta_n}.\eqno{(9)}
$$
Note the association of anomalous dimension and angle in
Equation (9), and the fact that it is completely factored --
each angle is raised to the power of the corresponding
operator only, in contrast to Equation (8).
Transforming Equation (9) to a new geometry, as above,
reproduces the correct conformal covariance of the
three-point function (Equation (3)) if one readjusts each
vertex of the transformed triangle to be at distance $\eps$
from the edge of the system.

If we let $\phi_n$ be the unit operator, Equation (9)
describes the two-point function, by a triangle with a
fictitious point $x_3$.
The resulting expression is not independent of $x_3$ unless
$\Delta_l=\Delta_m$, thus establishing orthogonality.

One can express the prefactor of an arbitrary four-point
function $G$ as a product of four similar factors, by
considering the hyperbolic quadrilateral defined by the four
points taken at distance $\eps$ above the real axis, as
above.
The remaining factor in $G$ is a function $\Phi$ of the
cross-ratio $C$.
This quantity may also be expressed through hyperbolic
angles.
If, for instance, one draws the triangle defined by points
$x_1$, $x_2$, and $x_3$, $C$ is given by the ratio of the
angle at $x_3$ of this triangle to the angle at $x_3$ of the
quadrilateral.

Now consider a four-point function $G$ of operators $\phi$
degenerate at level two, i.e. $\phi_{(1,2)}$ or
$\phi_{(2,1)}$.
This condition implies that the dimension $\Delta$ of $\phi$
is an algebraic function of the central charge $c$, and that
the factor $\Phi$ is proportional to a hypergeometric
function, i.e. there is a factor in $G$ that satisfies a
hypergeometric equation [10].
Now the ratio of two independent solutions of a
hypergeometric equation defines the Schwarz function, which
maps the upper half-plane onto a triangle with curvilinear
sides [17, 18].
In the present case, the triangle is equiangular, with angle
$\pi\vert\Delta'\vert$, where $\Delta'$ is the dimension of
the operator $\phi'$ (i.e. $\phi_{(1,3)}$ or $\phi_{(3,1)}$)
appearing in the operator product expansion of $\phi$ with
itself.
Now one may reflect the triangle across any of its sides,
which corresponds to a reflection of $C$ across the real
axis.
Repeating this procedure gives rise to a set of curvilinear
triangles that may overlap, i.e. the inverse map is not
necessarily single-valued.
If we require single-valuedness, the triangles will tile
(or, in the hyperbolic cases -- see below, {\it tessellate})
a circular region [18].
This condition (for either choice of $\phi$) specifies
$\Delta'=\vert 1\vert$, $1=\pm1, \pm2, \pm3,\ldots\,\,$.
Since $\Delta'$ determines $\Delta$, which in turn fixes the
central charge, one arrives at the doubly-infinite discrete
series
$$
c=1-3\,{(l-1)^2\over l(l+1)}.\eqno{(10)}
$$
Equation (10) specifies a set of minimal models, including
the Gassian model $(c=1,\,l=1)$, the Ising model
$(c=1/2,\,1=2)$, critical percolation and dilute polymers
$(c=0,\,l=3)$ [19], dense polymers [20] and matrix models
[21--23] $(c=-2,\,\vert l\vert=\infty)$, and the Yang-Lee
edge singularity $(c=-22/5,\,l=-5)$ [24].
For $\vert l\vert>3$, the sum of the angles is less than
$\pi$, so the triangles are hyperbolic, and a tessellation
is produced.
The inverse map is an automorphic function of the
corresponding triangular group.
For $\vert l\vert=1$, the triangle reduces to a great
circle, and the group consists of one element.
Similarly, $\vert l\vert=2$ gives the (finite) dihedral
group $\left<2,2,2\right>$ [18].
Both of these cases correspond to spherical geometry, with
total angle greater than $\pi$, while $\vert l\vert=3$ gives
rise to a triangular lattice in flat space (total angle
$\pi$).
The inverse map is an automorphic function of the group so
defined in each case.

In summary, we have established several connections between
the conformal field theory of boundary operators and
two-dimensional hyperbolic geometry.
A new interpretation of the basic equation of conformal
covariance arises, we find a fully factored form for the
three-point function, and a doubly-infinite discrete series
of central charges with limit $c=-2$ is discovered.
A correspondence between the anomalous dimension and the
angle of certain hyperbolic figures emerges.

\bigskip
P. K. is grateful for the hospitality of the School of
Mathematics, Institute for Advanced Study, where part of
this work was performed.
I. V. acknowledges support from the Laboratory for Surface
Science and Technology, University of Maine.
We would like to thank I. Rivin for discussions and a
particularly useful suggestion, and C. Callan, J. L. Cardy,
K. Dawson, J. Distler, M. E. Fisher, C. Johnson, R. P.
Langlands, E. Lieb, A. Ludwig, B. Nienhuis, A. Papadopoulos,
J. H. H. Perk, I. Peschel, R. Rietman, P. Sarnak, T.
Spencer, F. Williams and A. B. Zamolodchikov for stimulating
conversations.

\bigskip
\references{25}

{\nspace{
\ref{1.}
P. Kleban, Phys. Rev. Lett. 67,
2799 (1991).

\ref{2.}
H. W. Diehl and S. Dietrich, Z. Phys. B42, 65 (1981)
[and Erratum 43, 281 (1981)].

\ref{3.}
J. L. Cardy, Nucl. Phys. B324, 581 (1989).

\ref{4.}
J. L. Cardy and D. C. Lewellen, Phys. Lett. B259, 274
(1991).

\ref{5.}
D. C. Lewellen, Nucl. Phys. B372, 654 (1992).

\ref{6.}
A. Ludwig and I. Affleck, Phys. Rev. Lett. 67, 161 (1991).

\ref{7.}
A. M. Polyakov, Pisma ZhETP 12, 538 (1970) [JETP Lett. 12,
381 (1970)].

\ref{8.}
A. F. Beardon,
\sl the Geometry of Discrete Groups,
\rm Graduate Texts in Mathematics 91 (Springer, Berlin,
1983).

\ref{9.}
J. L. Cardy, private communication.

\ref{10.}
A. A. Belavin, A. M. Polyakov and A. B. Zamolodchikov,
Nucl. Phys. B241, 333 (1984).

\ref{11.}
J. L. Cardy, J. Phys. A25, L201 (1992).

\ref{12.}
R. P. Langlands, C. Pichet, P. Oouliot, and Y. Saint-Aubin,
J. Stat. Phys. 67, 553 (1992).

\ref{13.}
R. P. Langlands, P. Pouliot, and Y. Saint-Aubin,
preprint.

\ref{14.}
D. C. Lewellen, private communication.

\ref{15.}
C. Callan and F. Wilczek, Nucl. Phys. B340, 366 (1990).

\ref{16.}
R. Rietman, B. Nienhuis and J. Oitmaa, preprint.

\ref{17.}
I. S. Gradshteyn and I. M. Ryzhik,
\sl Table of Integrals, Series, and Products,
\rm corrected ed. (Academic Press, Orlando, 1980).

\ref{18.}
M. Yoshida,
\sl Fuchsian Differential Equations,
\rm Aspects of Mathematics Vol. E11 (Vieweg, Braunschweig,
1987).

\ref{19.}
V. S. Dotsenko and V. A. Fateev, Nucl. Phys. B240, 312
(1984).

\ref{20.}
H. Saleur, Nucl. Phys. B382, 486 (1992).

\ref{21.}
E. Br\'ezin and V. A. Kozakov, Phys. Lett. B236, 144
(1990).

\ref{22.}
M. Douglas and S. H. Shenker, Nucl. Phys. B335, 635 (1990).

\ref{23.}
D. J. Gross and A. A. Migdal, Nucl. Phys. B340, 333 (1990).

\ref{24.}
J. L. Cardy, Phys. Rev. Lett. 54, 1354 (1985).

}}

\bye